 \documentclass[preprint]{aastex} 

\slugcomment{Submitted to ApJ Letters} 
 
\shorttitle{Broad H$\alpha$ Wings in IC~4997} 
\shortauthors{Lee and Hyung} 
 
\begin{document} 
 
\title{Broad H$\alpha$ Wing Formation in the Planetary Nebula IC~4997} 
 
\author{Hee-Won Lee and Siek Hyung} 
\affil{Korea Astronomy Observatory,\\ 
     61-1, Whaam-dong, Yusong-gu, \\ 
    Taejon, 305-348, Korea} 
\email{hwlee@hanul.issa.re.kr} 
 
\begin{abstract} 
The young and compact planetary nebula IC~4997 is known to exhibit  
very broad wings with a width exceeding $5000\ {\rm km\ s^{-1}}$  
around H$\alpha$. We propose that the broad wings are formed through 
Rayleigh-Raman scattering involving atomic hydrogen, by which 
Ly$\beta$ photons with a velocity width of a few $10^2\ {\rm km\ s^{-1}}$  
are converted to optical photons and fill the H$\alpha$ broad wing region. 
The conversion efficiency reaches 0.6 near the line center where
the scattering optical depth is much larger than 1 and rapidly decreases
in the far wings.
Assuming that close to the central star there exists an unresolved inner 
compact core of high density, $n_H\sim 10^{9-10}\ {\rm cm^{-3}}$, 
we use the photoionization code `CLOUDY' to show that sufficient Ly$\beta$ 
photons for scattering are produced. Using a top-hat
incident profile for the Ly$\beta$ flux and a scattering region with 
a H~I column density $N_{HI}=2\times 10^{20}\ {\rm cm^{-2}}$ and a 
substantial covering factor, we perform a profile fitting analysis 
to obtain a satisfactory fit to the observed flux. 
We briefly discuss the astrophysical implications of the Rayleigh-Raman 
processes in planetary nebulae and other emission objects. 
 
\end{abstract} 
\keywords{planetary nebulae --- planetary nebulae : 
individual IC~4997 --- radiative transfer --- scattering --- profile} 
 
\section{Introduction} 
 
With the advent of the Hubble Space Telescope and 
large optical telescopes, planetary nebulae (PNe) came to be known to possess 
complex structures including multiple shells and collimated outflows, of which 
the origins still remain puzzling (e.g. Gurzadyan 1997). In addition,  
PNe display various nebular morphology ranging from elliptical to spherical, 
bipolar (or of butterfly shape), point-symmetrical and irregular. However, the
physical explanation for the morphological diversity is not found
(e.g. Corradi \& Schwarz 1995). 
 
Noting that symbiotic stars, widely known to be binary systems of a cool 
giant and a hot white dwarf or a main sequence star, exhibit  
bipolar emission nebulae, Soker (1998) advocated the view that the bipolar  
morphology in planetary nebulae is attributed to the binarity in the central  
star system. Evolutionary links between symbiotic 
stars and PNe have been also proposed by a number of 
researchers (e.g. Corradi 1995). Currently one of the most controversial 
issues is the existence of an accretion disk in symbiotic stars and bipolar 
PNe. According to the recent SPH computations by  
Mastrodemos \& Morris (1998), in a detached binary system consisting of  
a mass losing giant and a companion star, the slow and dusty stellar 
wind from the giant may induce the formation of an accretion disk 
around the companion star (see also Theuns et al. 1996). 
 
Circumstantial evidence supporting this scenario was provided  
by Lee \& Park (1999), who proposed that a disk type emission model 
can naturally explain the main characteristics of the profiles and 
the polarized fluxes of the Raman-scattered 6830 and 7088 \AA\ features 
in the symbiotic star RR Tel. 
The Raman scattering process by atomic hydrogen has been introduced by 
Schmid (1989), who identified the 6830 and 7088\ \AA\ bands 
in symbiotic stars as the Raman scattered O~VI 1032, 1038 doublet.  
A thick H~I component around the giant companion in a symbiotic star 
is believed to act as the Raman-scatterer which converts the UV continuum 
around Ly$\beta$ into the optical continuum. 
 
The young and compact PN IC~4997 is particularly interesting in that it  
also displays a hint of bipolar morphology, as is revealed in the radio 
image presented by Miranda \& Torrelles (1998).
Miranda et al. (1996) presented spectroscopy of IC~4997 which shows 
very broad wings around H$\alpha$ with FWHM $5375\ {\rm km\ s^{-1}}$. 
IC~4997 is not the only PN showing the broad H$\alpha$ wings.  
The bipolar planetary nebula M~2-9 is also known to exhibit 
extremely broad wings with width $\sim 10^4\ {\rm km\ s^{-1}}$ in H$\alpha$ 
(Balick 1989, see also Torres-Peimbert \& Arrieta 1998).  
 
Patriarchi and Perinotto (1991) used the IUE data of 159 planetary nebulae 
to find that about 60 per cent of central stars have a stellar wind 
with a measured edge velocity ranging 600 to $3500\ {\rm km\ s^{-1}}$.  
Feibelman (1982) presented IUE spectroscopy  of IC~4997  and found no 
evidence of P-Cygni profile in C~IV 1550 doublet, which strongly implies 
the absence of strong stellar winds around the central star. This makes 
it difficult to attribute the broad wing around H$\alpha$ in IC~4997 
to the kinematics associated with the fast stellar wind.   
 
Electron scattering has been often invoked to explain the wing 
formation in luminous blue variables (Bernat \& Lambert 1978). 
However, it does not appear to be the case in IC 4997 since 
the Thomson scattering cross section is independent 
of wavelength and therefore broad wings should be present around other 
emission lines. Furthermore, in a planetary nebula, 
it is difficult to incorporate an electron scattering component  
with the electron temperature $T_e > 10^6\ {\rm K}$ and a sufficient  
scattering optical depth. 
 
An interesting observation of IC~4997 is provided by Altschuler et al. (1986), 
who reported the detection of H~I in IC~4997 by observing the 21 cm absorption 
dip (see also Schneider et al. 1987). With a large amount of atomic hydrogen, 
the broad wings around H$\alpha$ can be formed by Rayleigh-Raman  
scattering, where the incident radiation in the 
vicinity of Ly$\beta$ converted to the radiation around H$\alpha$ with a 
width broadened by a factor of $\lambda_{H\alpha}/\lambda_{Ly\beta} = 6.4$ 
due to the incoherency of the scattering process.  In this Letter, we  
investigate the broad H$\alpha$ 
wing formation through the Rayleigh-Raman scattering process and discuss the 
implications on the astrophysics of bipolar planetary nebulae. 
 
\section{Model} 
 
\subsection{Wing Formation from the Rayleigh-Raman Process} 
 
In this work, we adopt a simple model to investigate the wing formation  
through the Rayleigh-Raman process. We assume that the 
scattering region is in the form of a finite slab  
with the column density $N_{HI}$ with the solid angle $\Omega_s$ with 
respect to the radiation source, which coincides with the central star 
of the planetary nebula. We discuss the Ly$\beta$ line radiation source 
in the next subsection and now we assume that the scattering region
is illuminated by a far UV radiation source with intensity $I_\lambda$.
 
A UV photon around Ly$\beta$ is incident on a hydrogen atom in the ground 
state $1s$, which subsequently de-excites either to the $2s$ state 
re-emitting an optical photon around H$\alpha$ (Raman scattering) or to 
the $1s$ state resulting in an outgoing UV photon with the same frequency 
(Rayleigh scattering). Depending on the scattering optical depth, 
a UV photon may suffer a number of Rayleigh scatterings followed by a 
Raman scattering and escape the scattering region, because the $2s$ state 
is usually populated negligibly compared with the ground state. 
 
The incident wavelength  
$\lambda_i$ and the wavelength $\lambda_o$ of the Raman-scattered radiation  
are related by the energy conservation, 
$\lambda_i^{-1} = \lambda_o^{-1} + \lambda_{Ly\alpha}^{-1}, $ 
which yields the conversion of the wavelength interval  
\begin{equation} 
\Delta\lambda_o/\lambda_o= (\lambda_o/\lambda_i) 
(\Delta\lambda_i/\lambda_i). 
\end{equation}  
Because of the factor $\lambda_o/\lambda_i$, UV photons 
with a width $\Delta V_i$ centered at Ly$\beta$ are spread around  
H$\alpha$ with a broadened width $\Delta V_o = 6.4 \Delta V_i$. 
 
The number of UV photons incident on the scattering region per unit time 
per unit wavelength interval is given by 
$I_{\lambda_i}(T_*) \pi R_*^2\Omega_s /(hc/\lambda_i)$. 
We introduce the parameter $C_R(\lambda_i)$, which is the conversion factor  
from UV to optical through the Rayleigh-Raman scattering process.
The conversion factor $C_R(\lambda_i)$ depends sensitively on the wavelength 
and the column density  as well as the detailed geometry of the scattering 
region, and can be computed using a Monte Carlo technique
(see Lee \& Yun 1998). 
 
The number of Raman-scattered photons coming out from  
the region per unit time per unit wavelength interval is given by 
\begin{equation} 
N_{H\alpha}=[I_{\lambda_i}(T_*) \pi R_*^2\Omega_s /(hc/\lambda)]C_R(\lambda) 
\lambda_i^2/\lambda_o^2. 
\end{equation} 
Therefore, the Raman-scattered flux is 
\begin{eqnarray} 
F^{Ram}(\lambda) &=& N_{H\alpha}(hc/\lambda)/r^2 \nonumber \\ 
&=& F^{UV}(\lambda_i)\times C_{eff} 
\end{eqnarray} 
where $F^{UV}(\lambda_i)\equiv [I_{\lambda_i}\pi R_*^2/ r^2]$ 
is the flux that would be observed from the radiation source with
a size $R_*$ located at a distance $r$  and 
$C_{eff}\equiv 
[\Omega_s C_R(\lambda_i)\lambda_i^3/\lambda^3]$ is the effective Raman 
conversion efficiency.

\subsection{Conversion Efficiency} 
In Fig.~1, we present the Raman conversion efficiency $C_R(\lambda)$ for 
H~I column densities $N_{HI}=10^{19}, 10^{20}$, and $10^{21}\ {\rm cm^{-2}}$, 
using the Monte Carlo code developed by Lee \& Yun (1998). 
In the optically thin regime, the single scattering approximation holds so  
that 
\begin{equation} 
C_R(\lambda) = \tau_{tot} (\sigma_{Ram}/\sigma_{tot}). 
\end{equation} 
In the optically thick limit, the conversion rate is investigated by 
Lee \& Lee (1997), who proposed an empirical relation 
\begin{equation} 
C_R(\lambda)=\sum_n r_{\sigma}f(n)/\sum_n[(1-r_{\sigma})\beta(n)+r_{\sigma}] 
f(n), 
\end{equation} 
where $r_{\sigma}\equiv \sigma_{Ram}/\sigma_{tot}$. Here,  
$\beta(n) \simeq {1\over2}e^{-\sqrt{n}}$ is the escape probability from the 
$n$th scattering site, and $f(n)$ is the photon number flux scattered no 
less than $n$ times which is obtained recursively by 
\begin{equation} 
f(n+1)=(1-r_{\sigma})[1-\beta(n)]f(n). 
\end{equation} 
A direct substitution with $r_{\sigma}=0.11$ near the line center 
yields $C_R(\lambda)=0.6$, which is in excellent agreement with the 
result shown in Fig.~1.

\subsection{Photoionization Structures of the Emission Nebula}

In this subsection we discuss the Ly$\beta$ flux that may enter the
scattering region.
Hyung et al. (1994) performed a photoionization computation to conclude
that the observed emission lines are reproduced approximately using
a model consisting of a thin shell of density $n\sim 10^7\ {\rm cm^{-3}}$
surrounded by a much larger shell with a much lower density $\sim 10^4\
{\rm cm^{-3}}$. They also noted that the N~III] $\lambda1754/\lambda1749$ 
ratio may indicate the existence of a much denser region with 
$n\ge 10^9\ {\rm cm^{-3}}$ (e.g. see Czyzak et al. 1986). 

We use the photoionization code `CLOUDY 90.05' developed by 
Ferland (1996) to obtain the Ly$\beta$ flux that is expected around
a central star of a PN. We fix the temperature and radius of the central star
to be $T_*=6\times 10^4\ {\rm K}$, $R_*=R_\odot$ and compute the hydrogen 
emission line fluxes for densities $n=10^{6}-10^{10}\ {\rm cm^{-3}}$.
According to Hyung et al. (1994) the observed H$\alpha$ luminosity is
$4\times 10^{35}\ {\rm erg\ s^{-1}}$ assuming the distance $d=2.4\ {\rm kpc}$
to IC~4997 and the interstellar extinction parameter 
$C$ =  log I(H$\beta$)/F(H$\beta$) = 0.8
where the observed flux F(H$\beta$) in H$\beta$ = 2.95$\times 10^{-11}$
ergs~s$^{-1}$~cm$^{-2}$.

In Table~1, we show the result. This shows that the observed H$\alpha$ flux
is obtained from a region of density $n\sim 10^7\ {\rm cm^{-3}}$, for
which we get a much smaller Ly$\beta$ flux. However, when $n\sim
10^{9-10}\ {\rm cm^{-3}}$, a much larger Ly$\beta$ flux is obtained.
Therefore, if there exists an emission region of density $n\sim 10^{9-10}\
{\rm cm^{-3}}$, we may have a sufficient number of Ly$\beta$ photons
to fill the H$\alpha$ wings. Furthermore, this region is plausibly located 
much closer to the central star than other emission regions are. 
Hence assuming that the distance is $\sim 0.1\ {\rm AU}$, we may expect 
that there is a high velocity dispersion of order $100\ {\rm km\ s^{-1}}$ 
in the region.

\section{Result}
 
In Fig.~2, by the dotted lines we show the H$\alpha$ wings expected from  
the Rayleigh-Raman processes. For the incident Ly$\beta$ flux we use 
$L_{Ly\beta} = 6\times 10^{35}\ {\rm erg\ s^{-1}}$, which is expected
to be obtained from an unresolved core of high density $n\sim 10^{9-10}\
{\rm cm^{-3}}$ near the central star.
We tested two types of profiles for
the incident Ly$\beta$ flux, that is, a top-hat profile with a width 
of $300\ {\rm km\ s^{-1}}$ and a Gaussian with the same velocity width.
We present the results for the best-fitting column density of 
the scattering region which is $N_{HI} =2\times 10^{20}\ {\rm cm^{-2}}$
for the incident top-hat profile and $N_{HI} =4\times 10^{20}\ {\rm cm^{-2}}$
for the Gaussian profile.  In this work, we concentrate only on the wing 
parts farther than $300 \ {\rm km\ s^{-1}}$ 
($\Delta \lambda > 6.5 \ {\rm \AA}$ around H$\alpha$) from the line center.  

Because the top-hat profile for the incident Ly$\beta$ provides an
excellent fit to the observed H$\alpha$ wing profile, 
we expect that other types of input profiles can also produce
a good fit 
except for some extreme cases. This is illustrated by another good fit
using a Gaussian profile shown in Fig.~2(b). 
However, we realize the much higher column
density, i.e. by a factor of 2, is required in the latter case.

In this work, the incident
Ly$\beta$ flux was computed with a highly simplified assumption. An 
introduction of a velocity field in the emission region may alter drastically
both the strength and the profile of the emergent line flux. Furthermore
the scattering region is complex and consists of a number of components
differing in column density and geometrical shape.
Therefore, a better fit can  be 
assured  from composite scattering region models and more complicated
radiative transfer, which are 
deferred to a future investigation. Since the extreme wings are formed by the  
scatterers located in the equatorial region, we  
predict stronger polarization in these regions than in the near center regions  
that are formed by scatterers distributed nearly spherically. 
  
\section{Discussion} 
 
The exact location of the scattering region with $N_{HI}= 10^{20}\ {\rm cm^{-2}}$  
is not well-constrained from this analysis.  
If the bipolar morphology has originated from a binary central 
star system as in symbiotic stars advocated by Soker (1998), 
 the binary companion of IC~4997  
may provide a scattering site in the form of 
an extended stellar atmosphere or a slow and probably dusty stellar wind. 
This situation is often met in symbiotic stars, where a slow and thick 
wind from a giant companion may act as a Raman scattering site. In this 
picture, the column density of $\sim 10^{20}\ {\rm cm^{-2}}$ is more 
plausibly attributed to the slow stellar wind. 
 
In the bipolar planetary nebula M2-9, Balick (1989) observed  the spectrum 
similar to that of the symbiotic star RR Tel. 
M2-9 might be the case of a binary central star system. Since there exists 
a thick circumstellar region in M2-9, the extremely wide  
H$\alpha$ wings with a width $11000\ {\rm km\ s^{-1}}$
are likely to be formed through the Raman process in a scattering region 
of column density $N_{HI}\sim 10^{21}\ {\rm cm^{-2}}$. 
 
Broad H$\alpha$ wings are found in a number of post-AGB stars 
according to a recent report by Van de Steene et al. (1999), who 
excluded several theoretical possibilities as the wing formation 
mechanism including Thomson scattering and Rayleigh 
scattering. Considering that in post-AGB stars much H~I can be found 
in the stellar wind region and/or in a dense torus surrounding the central 
star, it is very convincing that the broad H$\alpha$ wings in these systems 
are also attributable to Rayleigh-Raman scattering. 
So far the Raman-scattering is unique to symbiotic stars and the only 
exception appears to be the young planetary nebula NGC~7027, from which a 
Raman-scattered He~II line has been identified (P\'equignot et al. 1997). 
From this study broad H$\alpha$ wings are expected to be found in 
a hot and dense H~II region surrounded by an H~I region, where the
conditions may be met in objects including symbiotic stars, 
supernova remnants, and galactic supershells.
 
Since the Rayleigh-Raman scattering provides a strong constraint on the amount 
of H~I in the nebula, it can be an important tool in studying the mass loss 
process in the late stage of AGB evolution. Through high quality spectra 
around Balmer emission lines, the physical parameters of the central star 
can be estimated, which is complementary to the information obtained by 
other methods such as the Zanstra method. We also note that spectropolarimetry 
can provide more detailed structural information as mentioned in section 3.  
 
\acknowledgments 
 
HWL is grateful to Chul-Sung Choi and Hwankyung Sung for helpful comments and 
discussion. He also thanks the hospitality during his visit to the 
Korea Institute for Advanced Study. This  research was supported in part  
by Star Research Grant through No. Star~99-2-500-00  to the  
KAO/BOAO sponsored by the Korea Ministry of Science and Technology.

%\appendix 
% 
%\section{Appendicial material} 
% 

\clearpage
\input{epsf}
\begin{figure}
   \epsscale{0.5}
\centerline{\epsfxsize=5.5in\epsfysize=3.9in\epsfbox{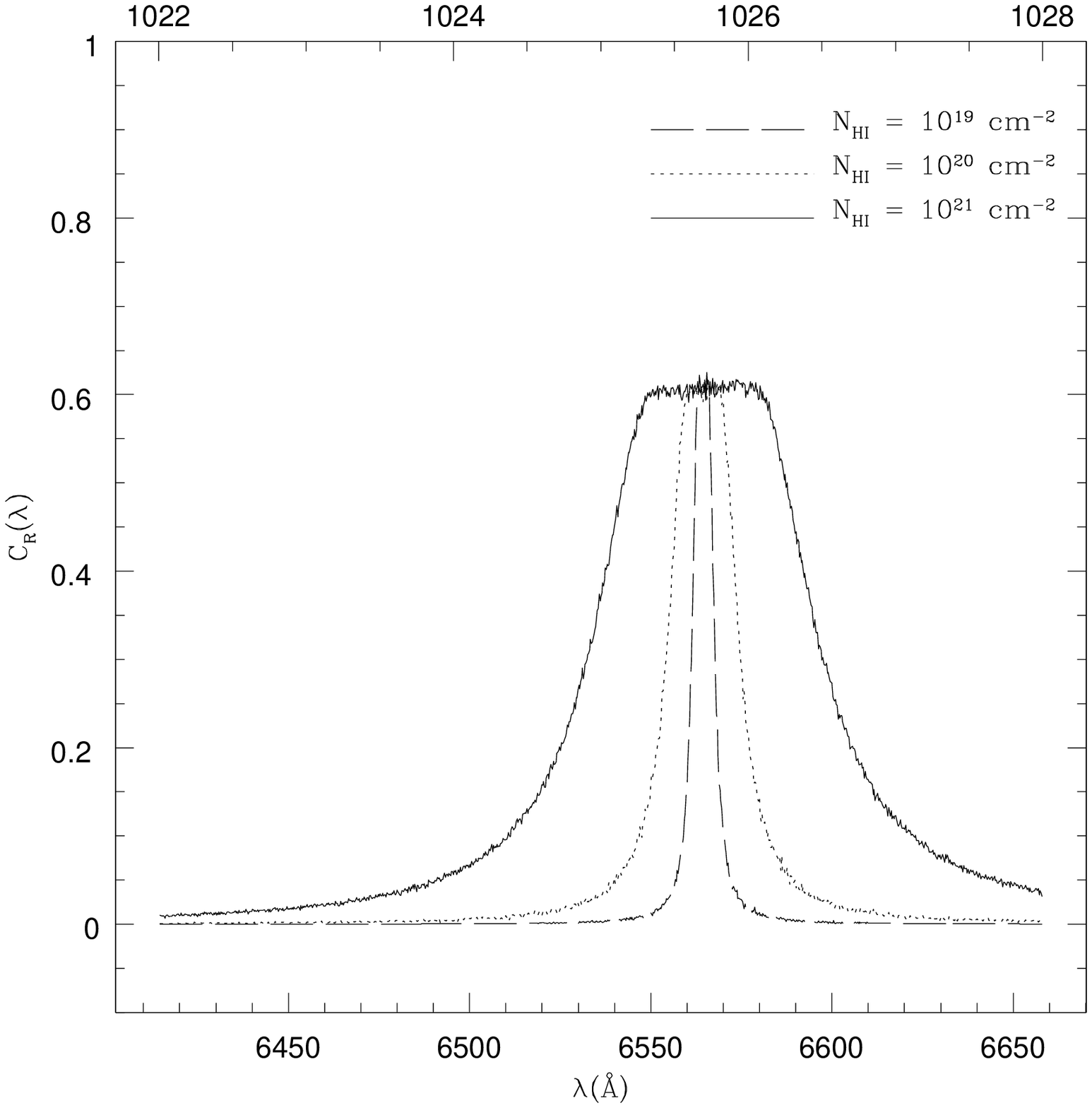}}
\caption[fig1_ic4.ps]{Raman conversion efficiency $C_R(\lambda)$ for 
H~I column densities $N_{HI}=10^{19}$ (solid line), $10^{20}$ (dotted line),  
and $10^{21}\ {\rm cm^{-2}}$ (long dashed line) obtained using a Monte Carlo  
technique. The horizontal (lower) axis is the  
outgoing wavelength around H$\alpha$ and in the upper part is shown the 
incident wavelength around Ly$\beta$.  
} 
\end{figure}
\begin{figure}
   \epsscale{0.5}
\centerline{\epsfxsize=5.5in\epsfysize=3.9in\epsfbox{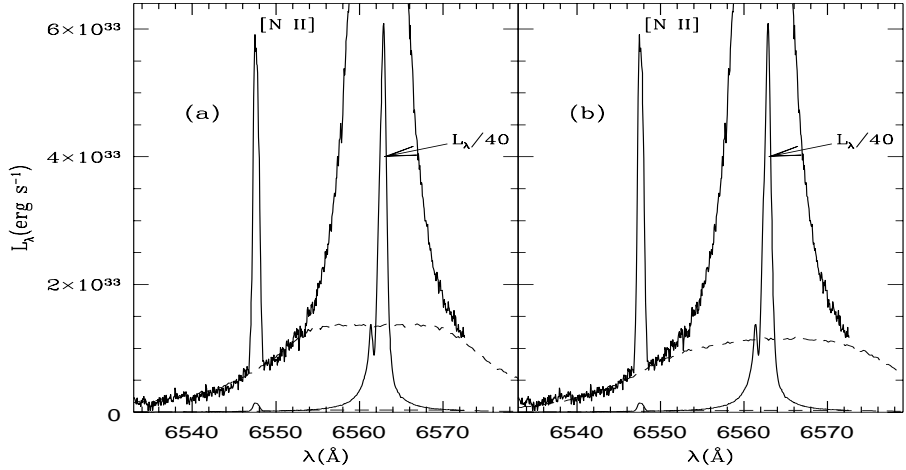}}
\figcaption[fig2_ic4.ps]{Observed flux around H$\alpha$ and the wings 
produced numerically in this work. 
The flux, represented by the solid line, was observed on  
August 31, 1991 UT, and is corrected for atmospheric and interstellar  
extinctions ($C = 0.8$). The vertical axis represents the specific luminosity
with the assumed distance $2.4\ {\rm kpc}$ to IC~4997. By the dotted lines
we show the wings produced via Rayleigh-Raman scattering. We assumed that the 
total incident Ly$\beta$ flux $L_{Ly\beta}=6\times 10^{35}\ {\rm erg\ s^{-1}}$.
In panel (a), the incident Ly$\beta$ has a top-hat profile with a width
$300\ {\rm km\ s^{-1}}$ and the H~I column density of the scattering region
$N_{HI} = 2\times 10^{20}\ {\rm cm^{-2}}$. In panel (b), the incident profile
is a Gaussian with the same width and the scattering region has $N_{HI}=
4\times 10^{20}\ {\rm cm^{-2}}$.
We also illustrate the flux multiplied by 1/40 to show the 
whole H$\alpha$ profile. 
}             
\end{figure}

\clearpage

\begin{deluxetable}{crrrrr}
\footnotesize
\tablecaption{H~I emission line fluxes from a plasma photoionized by a 
PN central star with $T_*=6\times 10^4\ {\rm K}$, and $R_*=R_\odot$.}
\tablewidth{0pt}
\tablehead{
\colhead{ $\log n$\tablenotemark{a} }&
\colhead{ $\log F_{Ly\alpha}$}  &
\colhead{ $\log F_{Ly\beta}$ }&
\colhead{ $\log F_{Ly\gamma}$ }&
\colhead{ $\log F_{H\alpha}$ }&
\colhead{ $\log F_{H\beta}$}
}
\startdata
10   & 36.849      &  35.687     & 35.154     & 35.382    & 34.988   \\
 9   & 36.944      &  35.233     & 34.544     & 35.874    & 35.451   \\
 8   & 36.957      &  34.025     & 33.135     & 35.957    & 34.991   \\
 7   & 36.961      &  33.268     & 33.043     & 35.890    & 35.318   \\
 6   & 36.953      &  33.146     & 33.049     & 35.875    & 35.380   \\
\enddata
\tablenotetext{a}{All quantities are measured in cgs units.}
\tablecomments{The photoionization code `CLOUDY 90.05' (Ferland 1996) was
used to obtain the result.}
\end{deluxetable}

\end{document}